%% file: 4LeptonsPaper.tex
\newcommand*{\ATLASLATEXPATH}{latex/}
\documentclass[UKenglish,texlive=2011,cernpreprint,txfonts,block=none]{\ATLASLATEXPATH atlasdoc}

\usepackage{\ATLASLATEXPATH atlaspackage}
\usepackage{\ATLASLATEXPATH atlasbiblatex}

\usepackage{\ATLASLATEXPATH atlascontribute}

\usepackage{\ATLASLATEXPATH atlasphysics}

\graphicspath{{logos/}{figures/}}

\usepackage{placeins}

\input{4LeptonsPaper-metadata}

\hypersetup{pdftitle={ATLAS draft},pdfauthor={The ATLAS Collaboration}}

\def\progname#1{{\textsc #1}}

\newcommand\MZZ{\ensuremath{m_{4\ell}}}

\newcommand{\xsecfid}{\ensuremath{\sigma_{4l}^\mathrm{fid}}}
\newcommand{\Nobs}{N_\mathrm{obs}}

\begin{document}
\nolinenumbers

\maketitle


\AtlasContributor{K.~Bachas}{background estimate using fake factor method, paper/note edit.}
\AtlasContributor{T.~Dai}{MC sample production.}
\AtlasContributor{H.~Feng}{MC sample production and theoretical calculation.}
\AtlasContributor{J.~Firmino~da~Costa}{background estimate using simultaneous fit, paper/note edit.}
\AtlasContributor{S.~Hassani}{cross section calculation using likelihood fit, paper/note edit.}
\AtlasContributor{S.~Hou}{supervisor for B.~Li.}
\AtlasContributor{J.F.~Laporte}{likelihood fit code for cross section calculation.}
\AtlasContributor{B.~Li}{theoretical calculation, signal acceptance and efficiency study, systematic uncertainty study, background estimate using fake factor method, cross section calculation using error matrix, paper/note edit.}
\AtlasContributor{J.~Liu}{background estimate using fake factor method}
\AtlasContributor{N.~Lu}{BDT method for $gg$ fit, paper/note edit.}
\AtlasContributor{N.~Orlando}{MEM method for $gg$ fit, paper/note edit.}
\AtlasContributor{C.~Petridou}{paper/note edit.}
\AtlasContributor{Y.~Wu}{unfolding and $gg$ fit, paper/note edit.}
\AtlasContributor{L.~Xu}{MC sample production, cross section calculation using error matrix.}
\AtlasContributor{Z.~Zhao}{supervisor for B.~Li and L.~Xu.}
\AtlasContributor{B.~Zhou}{supervisor for H.~Feng, N.~Lu and L.~Xu, paper/note edit.}  

\renewcommand{\figurename}{Figure}
\input{texfiles/Introduction.tex}
\input{texfiles/Detector_Section.tex}
\input{texfiles/Theory_Section.tex}

\input{texfiles/EventSelection.tex}
\input{texfiles/Background_Section.tex}
\input{texfiles/CrossSection_Section.tex}

\input{texfiles/Systematic_Section.tex}

\input{texfiles/Results_section.tex}
\input{texfiles/Unfolding_Section.tex}

\input{texfiles/ggExtraction_Section.tex}
\FloatBarrier
\input{texfiles/SummaryConclusions_Section.tex}

\input{acknowledgements/Acknowledgements.tex}

%

\printbibliography
\newpage \input{atlas_authlist}

\clearpage

\clearpage

%


\end{document}

%% file: 4LeptonsPaper-metadata.tex
\AtlasTitle{\input{texfiles/Title.tex}}
\author{The ATLAS Collaboration}

\AtlasRefCode{STDM-2014-15}

 \PreprintIdNumber{CERN-PH-EP-2015-220}

\AtlasJournal{Phys.\ Lett.\ B.}

\AtlasAbstract{%
\input{texfiles/Abstract.tex}
}

\AtlasCoverSupportingNote{ATL-COM-PHYS-2014-1357}{https://cds.cern.ch/record/1955922}
\AtlasCoverCommentsDeadline{25th September 2015}

\AtlasCoverAnalysisTeam{
  K.~Bachas, T.~Dai, H.~Feng,
  J.~Firmino~da~Costa, S.~Hassani, S.~Hou, J.F.~Laporte, B.~Li, J.~Liu, N.~Lu,
  J.~Luo,
  N.~Orlando, C.~Petridou, Y.~Wu, L.~Xu, Z.~Zhao, B.~Zhou
}

\AtlasCoverEdBoardMember{
Kevin Black, Daniel Fournier (chair), Yanyan Gao, Sasha Glazov (chair)
}

\AtlasCoverEgroupEditors{atlas-stdm-2014-15-editors@cern.ch}

\AtlasCoverEgroupEdBoard{atlas-stdm-2014-15-editorial-board@cern.ch}

%% file: texfiles/Title.tex
Measurements of four-lepton production in $pp$ collisions at $\sqrt{s}=$~8~\TeV{} with the ATLAS detector

%% file: texfiles/Abstract.tex
The four-lepton ($4\ell$, $\ell = e, \mu$) production cross section is measured in the
mass range from 80 to 1\,000 \GeV{} using 20.3~fb$^{-1}$ of data in $pp$
collisions at
$\sqrt{s}=8$~\TeV{} collected with the ATLAS detector at the LHC.
The $4\ell$  events are produced in the decays of resonant $Z$ and Higgs bosons
and the non-resonant $ZZ$ continuum originating from $q\bar q$, $gg$, and
$qg$ initial states. 
A total of 476 signal candidate events are
observed with a background expectation of $26.2 \pm 3.6$ events,
enabling the measurement of the integrated cross section and the
differential cross section as a function of the invariant mass and
transverse momentum of the four-lepton system.

In the mass range above $180$~\GeV{}, assuming the theoretical
constraint on the $q\bar q$ production cross section calculated with
perturbative NNLO QCD and
NLO electroweak corrections, the signal
strength of the gluon-fusion component relative to its leading-order
prediction is determined to be $\mu_{gg}=2.4 \pm 1.0~({\rm
stat.}) \pm 0.5~({\rm syst.})\pm 0.8~({\rm theory})$.

%% file: texfiles/Introduction.tex
\section{Introduction}
\label{sec:introduction}

\begin{figure*}[hbt]
\centering
  \subcaptionbox{}{\includegraphics[width=0.23\linewidth]{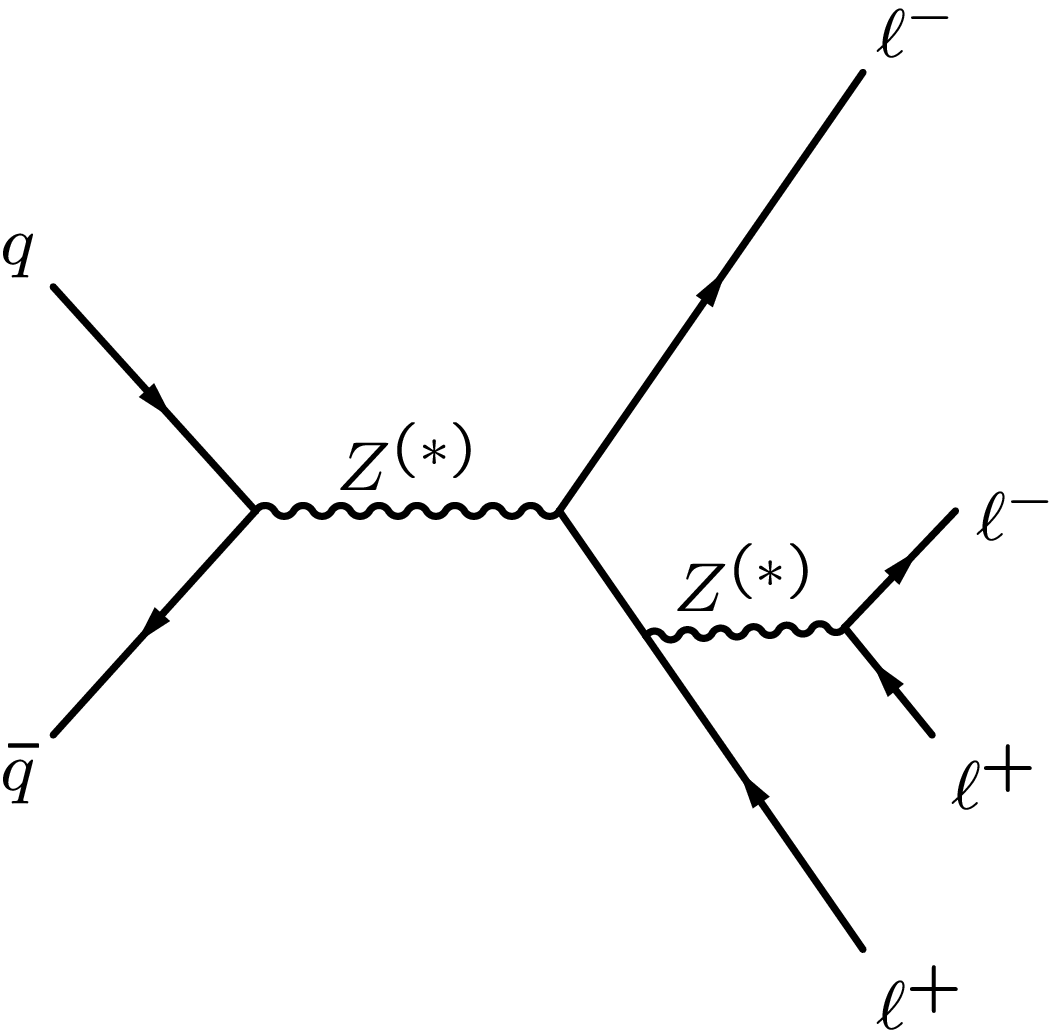}} \
  \subcaptionbox{}{\includegraphics[width=0.23\linewidth]{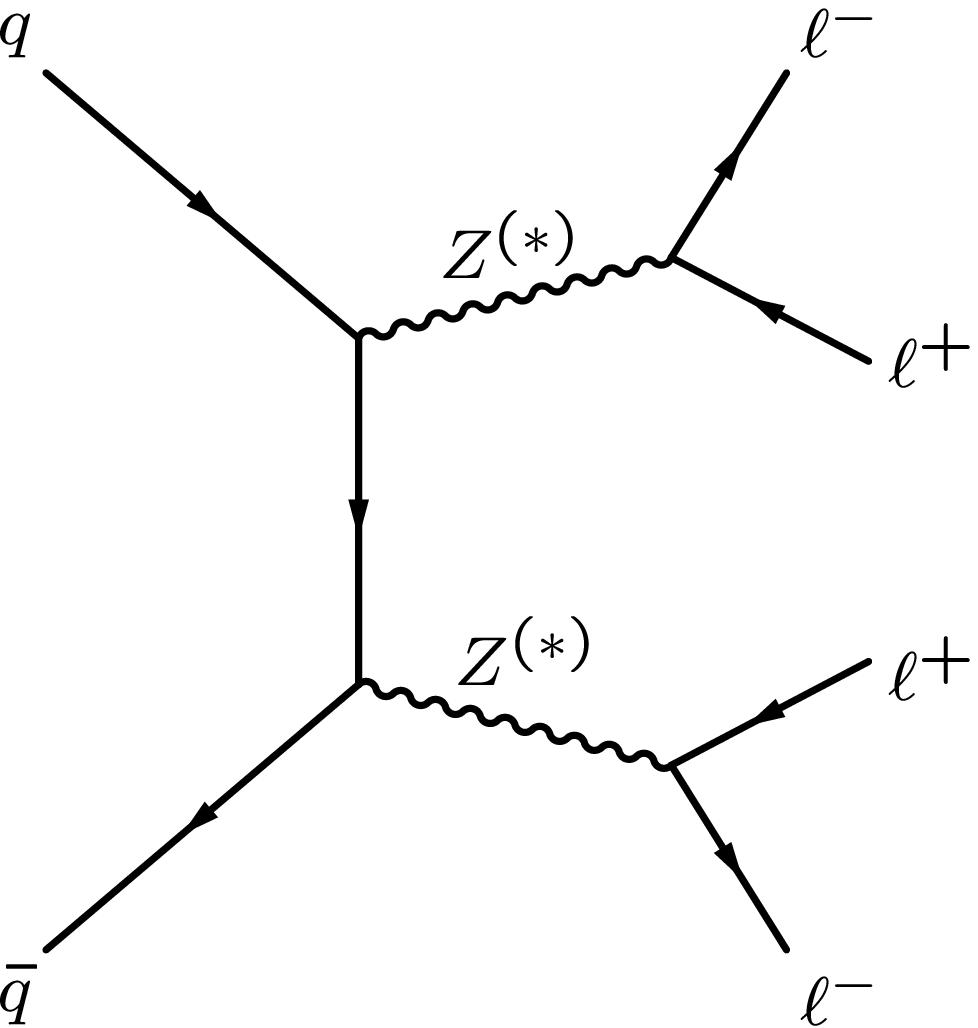}} \
  \subcaptionbox{}{\includegraphics[width=0.23\linewidth]{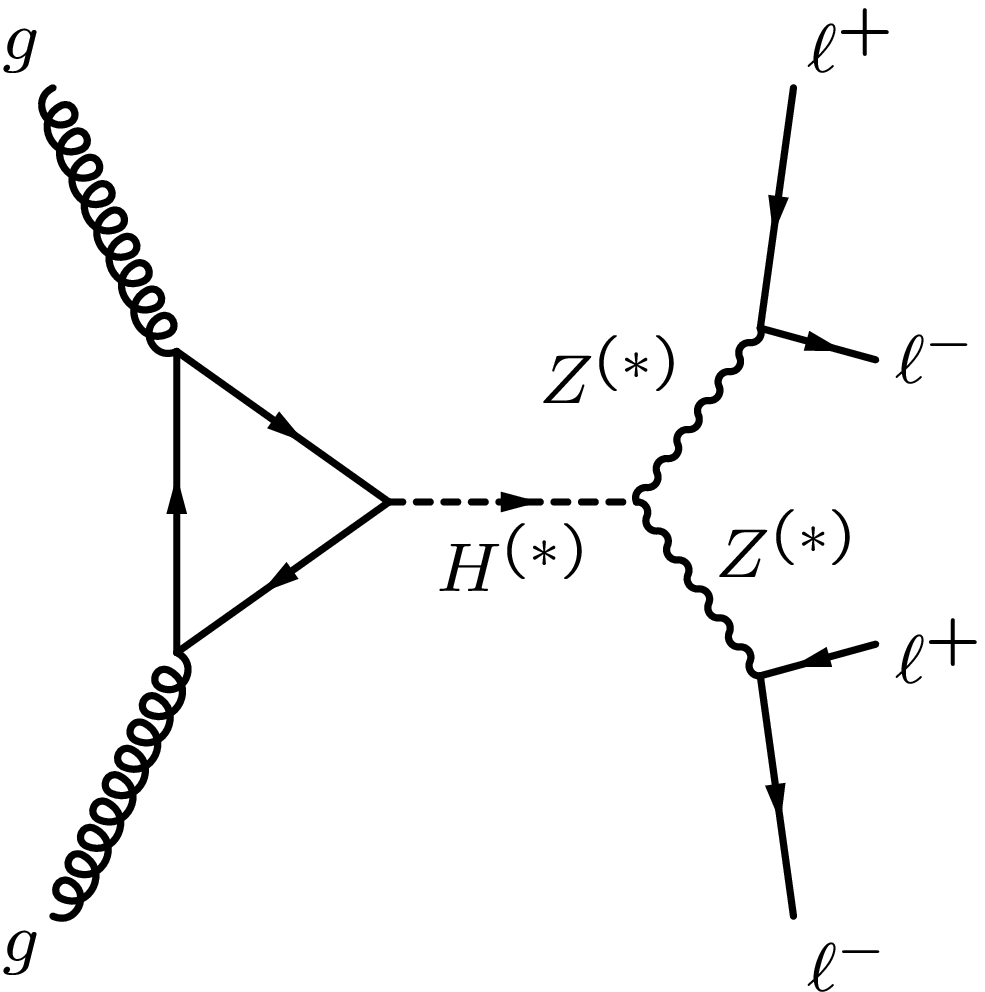}}  \
  \subcaptionbox{}{\includegraphics[width=0.23\linewidth]{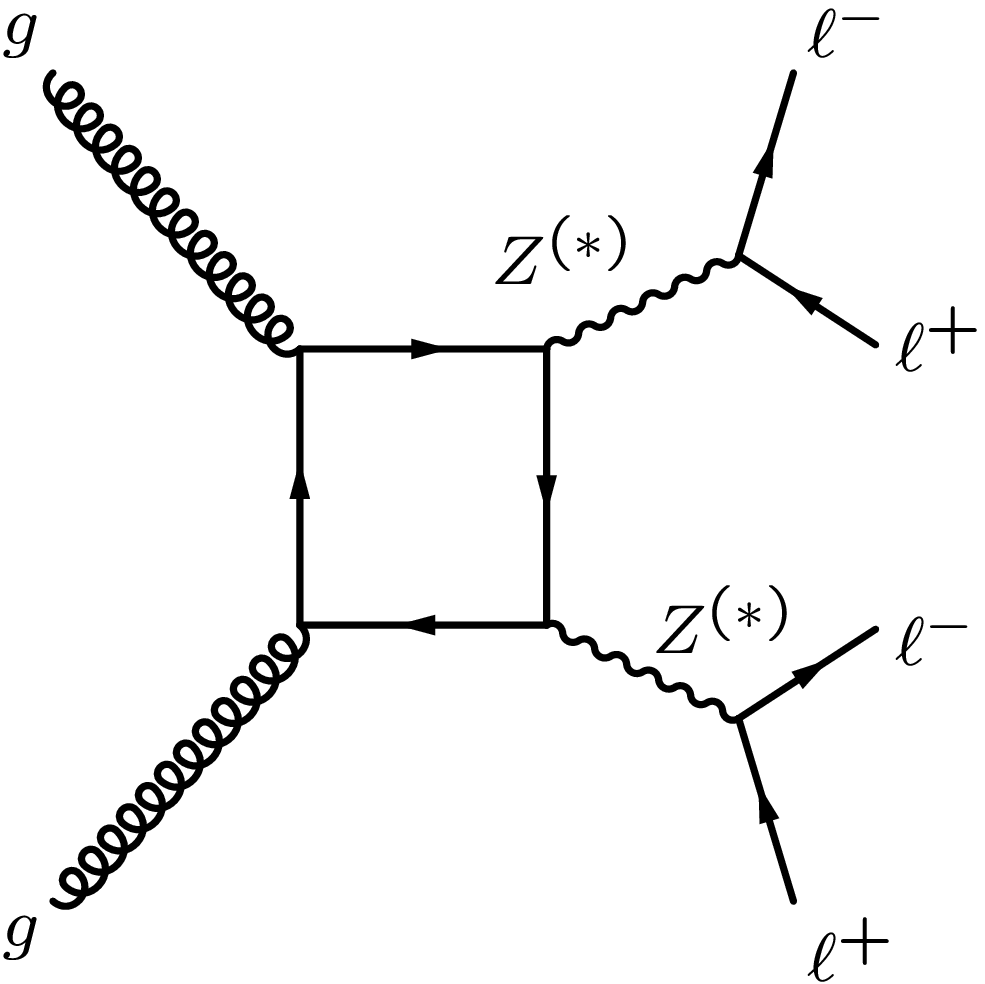}}
  \caption{
   The LO Feynman diagrams for the $q\bar{q}$- and $gg$-initiated production of 4$\ell$:
   (a) \mbox{$s$-channel}
      production of $q\bar{q} \rightarrow Z^{(*)} \rightarrow
      \ell^+\ell^-$ with associated radiative decays to an additional lepton pair;
   (b) \mbox{$t$-channel} production of $q\bar{q} \rightarrow
      Z^{(*)}Z^{(*)} \rightarrow 4\ell$;
   (c) Higgs-boson production through gluon fusion 
      $gg \rightarrow H^{(*)} \rightarrow ZZ^{(*)} \rightarrow 4\ell$;
   (d) non-resonant 4$\ell$ production through the quark-box diagram 
      $gg \rightarrow Z^{(*)}Z^{(*)} \rightarrow 4\ell$.
      The $Z^{(*)}$ notation stands for production of on- and off-shell $Z$ bosons 
      ($Z$ and $Z^{*}$) and production of off-shell photons ($\gamma^{*}$).
  }
  \label{fig:qqZZ-feynman}
\end{figure*}

This paper presents measurements of the production of  
four isolated charged-leptons 
in proton--proton collisions at a centre-of-mass energy of
$\sqrt{s}=8$~\TeV{} using 20.3~fb$^{-1}$ of data 
collected with the ATLAS detector at the LHC.  
For the four-lepton ($4\ell$, $\ell = e, \mu$)
production, both the integrated cross section
and the differential cross sections as functions of invariant mass ($m_{4\ell}$) 
and transverse momentum ($p_\mathrm{T}^{4\ell}$) of the $4\ell$ system,
are measured in a mass range 80 $< m_{4\ell} < $~1\,000~\GeV{}.
In addition, the $4\ell$ signal strength of
gluon fusion ({\it ggF}) production relative to its leading-order (LO)
QCD estimate is measured.  
These measurements test the validity of the
Standard Model (SM) through the interplay of QCD and electroweak
effects for different $4\ell$ production mechanisms
as described by the LO Feynman diagrams 
shown in Figure~\ref{fig:qqZZ-feynman}.

The $4\ell$ signal events come from the decays of resonant $Z$ and Higgs bosons
and the non-resonant {\it ZZ} continuum produced from $q\bar q$, $gg$, and
$qg$ initial states, which are briefly discussed below.

\begin{figure}[hbt]
\centering
\includegraphics[width=.95\linewidth]{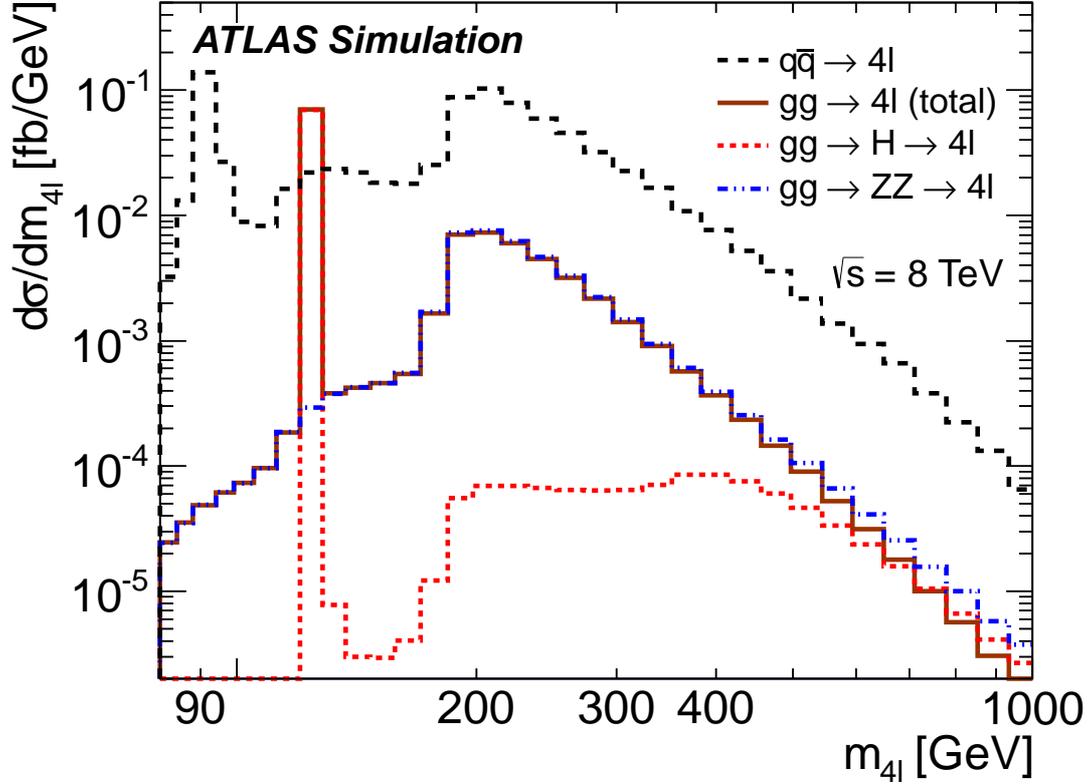}
  \caption{ The differential cross sections, $\mathrm{d}\sigma/d\MZZ{}$
            versus the invariant mass of the four leptons \MZZ{}, 
            calculated by \progname{MCFM}
  from the $q\bar{q}$ and $gg$ initial states at $\sqrt{s}$=8~\TeV{} for the
  $2e2\mu$ final state in the experimental fiducial phase space 
  (see Table~\ref{tab:FiducialRegion} for definition).
  The inclusive $gg \rightarrow 4\ell$ distribution is the sum of the
  $gg\rightarrow H\rightarrow 4\ell$ and the $gg\rightarrow ZZ
  \rightarrow 4\ell $, and interference terms.  The calculation of the
  $q\bar q \rightarrow 4\ell$ differential production cross section
  includes perturbative QCD corrections at NLO, while the
  distributions from the $gg$ initial state are calculated at LO.
  The NNLO $K$-factors 
  are applied to on-shell Higgs-boson production.
  }
  \label{fig:mcfm-mzz}
\end{figure}

\noindent $\bullet$ {\bf $ q\bar q $-initiated $4\ell$ production}\\
The tree-level diagrams for $q\bar q \rightarrow 4\ell$ production 
are shown in Figure~\ref{fig:qqZZ-feynman}(a) 
and Figure~\ref{fig:qqZZ-feynman}(b).
The cross section as a
function of $m_{4\ell}$ is shown in Figure~\ref{fig:mcfm-mzz}
(the dashed black histogram). 
The 4$\ell$ event production at
the $Z$ resonance occurs predominantly via the \mbox{$s$-channel}
diagram as shown in Figure~\ref{fig:qqZZ-feynman}(a), 
and was measured previously by the ATLAS and CMS
collaborations~\cite{atlas-singlez,cms-singlez}.  In the $4\ell$
invariant mass region above the $Z$ resonance the 4$\ell$ event
production mainly proceeds through the \mbox{$t$-channel} process as
shown in Figure~\ref{fig:qqZZ-feynman}(b). The cross section
significantly increases when both $Z$ bosons are produced on-shell,
resulting in a rise in the $m_{4\ell}$ spectrum around 180~\GeV.
In addition, a small portion of 
the $4\ell$ events with the $q\bar q$ initial state 
can be produced from the vector-boson scattering ({\it VBS}) process.
 
\noindent $\bullet$ {\bf $gg$-initiated $4\ell$ production}\\
The LO diagrams of the Higgs-boson production  
and non-resonant $4\ell$ production via {\it ggF}
are shown in Figure~\ref{fig:qqZZ-feynman}(c) 
and Figure~\ref{fig:qqZZ-feynman}(d), respectively.
The cross sections as a function of $m_{4\ell}$
are shown in Figure~\ref{fig:mcfm-mzz} (the colored histograms).
The features of the $4\ell$ events from the decays of Higgs-boson and
continuum $ZZ$ production via $ggF$ are described below.\\
$(1)$ The dominant Higgs-boson production mechanism is {\it ggF}.
Other Higgs-boson production mechanisms, vector-boson fusion ({\it VBF}), 
vector-boson associated production ({\it VH}), and top-pair associated production
($t\bar t${\it H}), 
contribute less than 15\% to the on-shell Higgs-boson decay to $ZZ^*$ event rate.
The on-shell Higgs-boson production
and decay leads to a narrow resonance around 125 GeV,
which has been a key signature in the Higgs-boson discovery by the ATLAS~\cite{atlas-higgs}
and CMS~\cite{cms-higgs} collaborations. 
The off-shell Higgs-boson
production has a large destructive interference with continuum {\it ZZ}
production from the {\it ggF} processes~\cite{gg2vv,Caola:2013yja,Goria:2011wa}.  
This effect can be observed in the high-mass tail of the distributions 
shown in Figure~\ref{fig:mcfm-mzz}, 
and has been used as a tool to constrain the total Higgs-boson width by
the ATLAS and CMS collaborations~\cite{atlas-hwidth-paper,cms-hwidth}.\\
$(2)$ The non-resonant $ZZ\rightarrow 4\ell$ production via {\it ggF}, 
includes the production of off-shell Higgs bosons and continuum
{\it ZZ} production as well as their interference.  
This process produces a sizeable number of $4\ell$ events in the 
$m_{4\ell} > 2\times m_{Z}$ mass region and dominates the total $gg$-initiated
$4\ell$ production.

Contributions from different processes have different strengths as a
function of $m_{4\ell}$ (Figure~\ref{fig:mcfm-mzz}) and $p_\mathrm{T}^{4\ell}$.
Therefore, differential $4\ell$ production cross
sections are measured separately as a function of $m_{4\ell}$ and $p_\mathrm{T}^{4\ell}$.
The measurement of the integrated cross section 
is first performed in the experimental fiducial phase space,
and then extended to a common phase space for three $4\ell$ channels:
$4e$, $4\mu$, and $2e2\mu$.
This common phase space is defined by
80 $< m_{4\ell} < 1\,000~\GeV$, 
$m_{\ell^+\ell^-} > 4$~\GeV, 
$p_\mathrm{T}^{Z_{1,2}} > 2\GeV$, 
and the presence of 
four leptons each with $p_\mathrm{T} > 5 \GeV$ and $|\eta| < 2.8$.

Currently, the gluon-fusion production is estimated
theoretically with only a LO QCD approximation 
for the $gg$ continuum production~\cite{Caola:2013yja,resummation}.
In this analysis
the mass range above $180$~\GeV{} is used to determine the signal
strength of the gluon-fusion component with respect to its LO
prediction.  This is done by fitting the observed $m_{4\ell}$ spectrum
using 
the next-to-next-to-leading-order (NNLO) QCD theoretical prediction,
corrected for next-to-leading-order (NLO) electroweak effects,
for the production originating from the
$q\bar{q}$ initial state.

%% file: texfiles/Detector_Section.tex
\section{The ATLAS detector}
\label{sec:Detector}

The ATLAS detector~\cite{bib:ATLASDetectorPaper}
has a cylindrical geometry\footnote{ATLAS uses a right-handed coordinate system with its origin at the nominal interaction point (IP) in the centre of the detector and the $z$-axis along the beam pipe. The $x$-axis points from the IP to the centre of the LHC ring, and the $y$-axis points upward. Cylindrical coordinates $(r,\phi)$ are used in the transverse plane, $\phi$ being the azimuthal angle around the beam pipe. The pseudorapidity is defined in terms of the polar angle $\theta$ as $\eta=-\ln\tan(\theta/2)$.}  
and consists of an inner tracking detector (ID) surrounded by a 2~T superconducting solenoid, 
electromagnetic and hadronic calorimeters, and a muon spectrometer (MS) with a toroidal magnetic field. 
The ID provides tracking for charged particles for $|\eta|<2.5$. 
It consists of silicon pixel and strip detectors surrounded by a straw tube tracker that 
also provides transition radiation measurements for electron identification. 
The electromagnetic and hadronic calorimeter system covers the pseudorapidity range $|\eta|< 4.9$. 
For $|\eta|<2.5$, the liquid-argon 
electromagnetic calorimeter is finely segmented and plays an important role 
in the electron identification.
The MS includes fast trigger chambers ($|\eta|<2.4$) 
and high-precision tracking chambers covering $|\eta|<2.7$.
A three-level trigger system selects events to be recorded for offline physics analysis. 

%% file: texfiles/Theory_Section.tex
\section{Signal and background simulation}
\label{sec:Theory}

The signal modelling for $q\bar q \rightarrow 4\ell$ production uses
the \progname{POWHEG-BOX} Monte Carlo (MC) program~\cite{Melia:2011tj,Alioli:2010xd,Nason:2004rx},
which includes perturbative QCD corrections at NLO.
The production through the $qg$ initial state is 
an NLO contribution to the $q \bar q$ process.
The CT10NLO~\cite{Lai:2010vv} set of parton distribution functions (PDFs),
with QCD renormalisation and factorisation scales
($\mu_\mathrm{R}, \mu_\mathrm{F})$ set to $m_{4\ell}$ are used to calculate the cross section
and generate the kinematic distributions.
The NNLO QCD~\cite{nnlocorr}
and the NLO electroweak (EW)~\cite{ewcorr} corrections
are applied to the NLO cross section calculated by \progname{POWHEG-BOX} 
as a function of the $4\ell$ mass for the kinematic region where both $Z$ bosons are produced on-shell.
Following the same approach as described in Ref.~\cite{atlas-hwidth-paper}, 
the $4\ell$ event distributions are re-weighted to
match those expected when using QCD scales of $m_{4\ell}/2$. 
This is done to unify the QCD scales 
used in simulation of the $q\bar q$ and the $gg$ processes.

The signal modelling of the on-shell Higgs-boson production via the {\it ggF} and {\it VBF} mechanisms
uses \progname{POWHEG-BOX} which provides calculations at NLO QCD,
with the CT10NLO PDFs and $\mu_\mathrm{R}, \mu_\mathrm{F} = m_{4\ell}$.
The Higgs-boson production via the {\it VH} and $t\bar t${\it H} mechanisms
is simulated with \progname{PYTHIA8}~\cite{PYTHIA8}. 
The NNLO QCD and NLO EW effects on the 
cross-section calculations for
on-shell Higgs-boson production 
are summarised in Ref.~\cite{yellowreport}.
The expected event yields of on-shell Higgs boson are normalised to 
the higher-order corrected cross sections. 

The non-resonant $4\ell$ signal production includes off-shell Higgs-boson production, continuum $ZZ$ production,
and their interference.
The LO \progname{MCFM} generator~\cite{Campbell:2013una}
is used to simulate the non-resonant {\it ggF} production,
with the CT10NNLO~\cite{ct10nnlo} set of PDFs
with QCD scales of $\mu_\mathrm{R}, \mu_\mathrm{F}$ set to $m_{4\ell}/2$;
while the LO \progname{MADGRAPH} generator~\cite{madgraph} 
is used to simulate non-resonant {\it VBF} and {\it VBS} production and their interference.
The NNLO QCD corrections are available for
off-shell Higgs-boson production~\cite{kfactor-ggzz}
and for the 
interference between off-shell Higgs bosons and $ZZ$ pairs from the $gg$ initial state~\cite{kfactor-ggzzinterference}.
However,
no higher-order corrections are available
for the continuum $gg \rightarrow ZZ$ process,
which dominates the $4\ell$ events from the $gg$ initial state
in the region outside the Higgs-boson resonance.
Therefore, the LO cross section is used
for the normalisation of the $4\ell$ events produced in 
gluon-fusion processes.

All the signal MC generators are interfaced to \progname{PYTHIA8} for parton shower
simulation, except \progname{MADGRAPH}, which is interfaced to \progname{PYTHIA6}~\cite{pythia6}.

Backgrounds in this analysis include
reconstructed $4\ell$ events from $Z$+jets, $t\bar t$, diboson 
({\it ZW}, {\it Z}$\gamma$ and double Drell-Yan), 
triboson {\it VVV} ({\it V=Z, W}), 
and {\it VH} ({\it H} $\rightarrow$ {\it VV}), and $Z$+top ($t\bar t$ and $t$) processes, which
are also simulated.  

The reducible background from $Z+\rm{jets}$ production, 
which includes light- and heavy-flavour contributions, is modelled 
using both \progname{SHERPA}~\cite{SHERPA} and \progname{ALPGEN}~\cite{mangano:2002ea}.
The $Z\rm{\gamma}$ process is simulated with  \progname{SHERPA}.
The $t\bar{t}$ background is modelled using \progname{POWHEG-BOX}.

Background events from {\it ZH} production, 
where $Z \rightarrow \ell \ell$ and $H \rightarrow VV$
({\it VV = WW} or {\it ZZ} with two leptons and two neutrinos or two leptons 
and two jets in the final state),
are simulated with \progname{PYTHIA8}.
The {\it ZW} and
the {\it tZ} processes are simulated with  
\progname{SHERPA} and \progname{MADGRAPH}, respectively.
The irreducible background from {\it VVV}
and $t \bar t${\it Z} is modelled with \progname{MADGRAPH}.
Finally, the double-Drell-Yan {\it ZZ} production is modelled with \progname{PYTHIA8}.

For background modelling the \progname{POWHEG-BOX} and \progname{MADGRAPH} 
generators are interfaced to \progname{PYTHIA6} 
for the parton shower,
hadronisation and underlying-event simulation. The \progname{ALPGEN}
generator is interfaced to \progname{HERWIG}~\cite{HERWIG} for the
parton shower and to \progname{JIMMY}~\cite{JIMMY} for the underlying
event simulation. \progname{SHERPA} uses built-in models for both
the parton shower and underlying-event description.

Both the signal and background MC events are 
simulated using the ATLAS detector
simulation~\cite{simuAtlas} based on the GEANT4~\cite{geant4} framework.
Additional $pp$ interactions in the same and nearby bunch crossings (pile-up)
are included in the simulations.
The MC samples are re-weighted to reproduce the distribution of
the mean number of interactions per bunch crossing observed in the data.

%% file: texfiles/EventSelection.tex
\section{Event reconstruction and selection}
\label{sec:EventSelection}

The following event selection criteria are applied to the events 
collected with a single-lepton or dilepton trigger.
The transverse momentum and transverse energy thresholds
of the single-muon and single-electron triggers are
24~\GeV{}. Two dimuon triggers are used, one with
symmetric thresholds at 13~\GeV{} and the other with
asymmetric thresholds at 18~\GeV{} and 8~\GeV{}. For
the dielectron trigger the symmetric thresholds are
12~\GeV{}. Furthermore, there is an electron--muon
trigger of thresholds at 12~\GeV{} (electron) and
8~\GeV{} (muon).

A primary vertex reconstructed from at least three 
tracks, each with $p_\mathrm{T} > 0.4$~\GeV, is required.
For events with more than one primary vertex, 
the vertex with the largest $\sum p_\mathrm{T}^2$ of the associated tracks is selected.

Electron candidates are reconstructed from a combination of a cluster of energy deposits 
in the electromagnetic calorimeter and a track in the ID. 
They are required to have $p_\mathrm{T} > 7$~\GeV\ and $|\eta| < 2.47$. 
Candidate electrons must satisfy a loose set of identification criteria based on a likelihood 
built from parameters characterising the shower shape and track association as described in Ref.~\cite{eID_Likelihood}.

Muon identification is performed according to several criteria based on the information from the ID,  
the MS, and the calorimeter.
The different types of reconstructed muons are:
a) Combined (CB), which is the combination of tracks reconstructed independently in the ID and MS;
b) Stand-Alone (SA), where the  muon  trajectory  is  reconstructed  only  in  the  MS;  
c) Segment-tagged (ST), where a track in the ID is associated with at least one local track segment 
in the MS;  
and d) Calorimeter-tagged  (CaloTag), 
where a track in the ID is identified as a muon if it is associated 
with a minimum ionizing particle's energy deposit in the calorimeter. 

The acceptance 
for both the CB and ST muons is $|\eta|<2.5$, 
while the SA muons are used to extend the $|\eta|$ acceptance 
to include the region from 2.5 to 2.7,
which is not covered by the ID. 
CaloTag muons are used in the rapidity range $|\eta|<0.1$ where there
is incomplete  MS coverage. 
All muon candidates are required to have $p_\mathrm{T} >6$~\GeV{}.

In order to reject electrons and muons from hadron decays, 
only isolated leptons are selected. 
Two isolation requirements are used, one for the ID and one for the calorimeter.
For the ID, the requirement is that the scalar sum of the transverse momenta, $\sum p_\mathrm{T}$, 
of all tracks inside a cone of
$\Delta R \equiv \sqrt{(\Delta\eta)^2 + (\Delta\phi)^2} = 0.2$ around the lepton, 
excluding the lepton itself, be less than 15\% of the lepton $p_\mathrm{T}$.
For the calorimeter,
the $\sum \ET$ deposited inside a cone of
$\Delta R = 0.2$ around the lepton, excluding the lepton itself and corrected 
for contributions from pile-up and, 
in the case of electrons, shower leakage, 
is required to be less than 30\% of the muon $p_\mathrm{T}$ (15\% for SA muons) 
and 20\% of the electron $E_\mathrm{T}$.

At the closest approach of a track to the primary vertex,
the ratio of the transverse impact parameter $d_0$ to its uncertainty, 
the $d_0$ significance, 
must be smaller than 3.5 (6.5) for muons (electrons) to 
further reject leptons from heavy-flavour decays. 
The longitudinal impact parameter, $|z_0|$, must be less than $10$ mm 
for electrons as well as muons (no vertex requirements are applied to SA muons).

Selection of lepton quadruplets is done separately in each of the channels 
$4\mu,2e2\mu,4e$, keeping only a single quadruplet per channel.
Candidate quadruplets are formed by selecting two opposite-sign, same-flavour 
lepton pairs ($\ell^+\ell^-$).
The two leading-$p_\mathrm{T}$ leptons of the quadruplet must 
have $p_\mathrm{T} > 20$ and 15~\GeV{}, respectively,
while the third lepton must have $p_\mathrm{T} > 10 ~(8)$~\GeV{} if it is an electron (muon).
The four leptons of a quadruplet are required to be separated from each other by 
$\Delta R >0.1~(0.2)$ for same (different) flavour. 
At most one SA or a CaloTag muon is allowed in each quadruplet.
The inclusion of final-state radiation to charged leptons follows
the same approach as described in Ref.~\cite{atlas-higgs4lRun1}. 
Each event is required to have the triggering lepton(s) matched 
to one or two of the selected leptons.
All the selected $4\ell$ events must 
lie in the $80 < m_{4\ell} < 1\,000$~\GeV{} range.

For each channel, the lepton pair with the mass closest to the $Z$-boson mass 
is selected as the leading dilepton pair and its invariant
mass, $m_{12}$, is required to be between 50 and 120~\GeV{}.
The sub-leading $\ell^+\ell^-$ pair with the largest invariant mass, $m_{34}$, 
among the remaining possible pairs, is selected in the invariant mass range 12 $ < m_{34} < 120$~\GeV.
In the $4e$ and $4\mu$ channels all possible $\ell^+\ell^-$ pairs are required to 
have $m_{\ell^+\ell^-} > 5$~\GeV{} to reject events containing $J/\psi\rightarrow \ell^+\ell^-$ decays.
The transverse momenta of the lepton pairs must be above 2~\GeV{}. 

%% file: texfiles/Background_Section.tex
\section{Background estimation}
\label{sec:BkgEstimation}

The dominant reducible 
backgrounds for this analysis are
from $Z+\rm{jets}$ and $t\bar{t}$ processes and are estimated from data.
Contributions from {\it ZW}, {\it Z}$\gamma$, {\it tZ} as well
as from the irreducible backgrounds from $t\bar{t}${\it Z}, {\it VVV}, {\it ZH} and 
double-DY processes are estimated from simulation.

The {\it Z}+{\it jets} and $t\bar{t}$ backgrounds are estimated using two different final states in data: 
$\ell\ell+\mu\mu$ and $\ell\ell+ ee$, where $\ell\ell$ $(\ell=e,\mu)$ is the leading-lepton pair.
The $\ell\ell+\mu\mu$ background arises from $Z+{\rm jets}$ and 
$t\bar{t}$ processes where the $Z+{\rm jets}$ contribution 
involves the associated production of a Z boson and heavy-flavour hadrons, which decay semileptonically, and 
a component arising from $Z$~+~light-flavour jets with subsequent $\pi$/\kaon~in-flight decays. 
The background for $\ell\ell+ee$ final states arises from associated
production of a Z boson with other objects namely jets 
misidentified as electrons, which can be light-flavour hadrons misidentified as electrons, 
photon conversions reconstructed as electrons, or 
electrons from semileptonic decays of heavy-flavour hadrons.

For both the $\ell\ell+\mu\mu$ and $\ell\ell+ee$ cases, 
the numbers of background events are estimated 
from a fit performed simultaneously to three 
mutually exclusive control regions, 
each of them providing information on one or more background components. 
The fit is based on the mass of the leading dilepton, $m_{12}$, which
peaks at the $Z$ mass for the $Z+{\rm jets}$ component and has a broad distribution 
for the $t\bar{t}$ component. 
The three control regions are fit simultaneously to extract the different components 
of the reducible background, using a profile likelihood approach 
where the input template shapes for $Z+{\rm jets}$ and $t\bar{t}$  
are obtained from simulation. 
The fitted yields in the control regions are extrapolated to the signal region 
using efficiencies, referred to as transfer factors, obtained from simulation.
Independent validation regions are used to check the extrapolations.

The three control regions for $\ell\ell+\mu\mu$ background
are defined based on the impact parameter significance and isolation variables 
of the sub-leading muon pair and are constructed as follows:
\begin{itemize}
\item A heavy-flavour-enriched control region where at least one of the muons 
in the second pair fails the impact parameter significance requirement while the isolation requirement is relaxed;
\item A light-flavour-enriched control region where at least one of the muons in the second pair fails 
the isolation requirement but passes the impact parameter significance cut;
\item A $t\bar{t}$-enriched region  where the leading lepton pair is made of opposite-sign 
and different-flavour leptons. For the muons of the second pair there is no charge requirement, 
the isolation cut is relaxed and the muons must not satisfy the impact parameter requirement.
\end{itemize}
A validation region to check the $\ell\ell+\mu\mu$ background extrapolation
is populated by both $Z+{\rm jets}$ and $t\bar{t}$. 
The leading lepton pair is required to fulfil the full selection criteria, 
while there is neither isolation nor impact parameter requirements on the sub-leading muon pair. 
This region is used to check the fit results and verify that the data and MC simulation agree.

The three control regions for $\ell\ell+ee$ background
are defined based on the impact parameter significance, isolation and electron identification requirements on the second electron pair. In all control regions at least one of the electrons in the second pair must not satisfy the identification criteria.
These regions are constructed as follows:
\begin{itemize}
\item A $Z+{\rm jets}$-enriched control region where at least one of the electrons of the second pair fails the track isolation and no calorimeter isolation is required;
\item An additional $Z+{\rm jets}$-enriched control region where no charge requirement is made on the electrons of the second pair, while at least one of these electrons fails the impact 
parameter selection and no calorimeter or track isolation is required;
\item A $t\bar{t}$-enriched region, where the leading lepton pair is selected from opposite-sign 
and different-flavour leptons. There is no charge requirement for the sub-leading electron pair. 
At least one of the electrons of the second pair fails the calorimeter isolation requirement 
and neither track isolation nor impact parameter requirements are applied.
\end{itemize}
A validation region to check the $\ell\ell+ee$ background extrapolation 
is defined by removing the calorimeter isolation and requiring that at 
least one electron in the sub-leading pair fails the electron identification. 
Each candidate in the pair is required to pass the impact parameter and the track 
isolation selections.  
this region is used to check the fit outcome 
and verify that the data and MC simulation agree.

The residual contributions from {\it ZZ} and {\it ZW} production in all control regions 
are estimated from simulation.
The purity of the {\it Z}+{\it jets} and $t\bar{t}$ backgrounds 
in the control regions is above $95\%$. 

In the validation regions, the post-fit MC predictions agree with the data within the statistical uncertainty. 

The major uncertainties for the fitted reducible background 
come from the number of events in the control regions 
followed by  the systematic uncertainty in the transfer
factors. 
The latter is evaluated from the difference in the selection efficiency determined in data and simulation 
in dedicated control regions using leptons
accompanying $Z \rightarrow \ell^+\ell^-$ candidates, where the leptons composing the $Z$-boson 
candidate are required to satisfy isolation and impact parameter
criteria. Events with four leptons are excluded. 
For the MC estimated background the systematic uncertainties 
mainly come from
theoretical cross-section uncertainties for different processes
and from luminosity uncertainties in normalisations.
The differential distributions for all background processes are taken from simulation.

\begin{table}[!htb]
\caption{
Numbers of expected background events for different processes and channels.
}
\centering
\begin{tabular}{cccc}
\toprule
Process            & $4e$ & $4\mu$  & $2e2\mu$ \\             
\midrule           
\ttbar\            & $0.45\pm0.24$     & $0.68\pm0.19$        & $1.3 \pm0.5 $        \\
$Z+{\rm  jets}$    & $0.6 \pm0.29$     & $5.3 \pm1.5 $        & $6.3 \pm1.4  $        \\
Diboson            & $1.25\pm0.18$     & $0.83\pm0.18$        & $2.84\pm0.34$        \\ 
Triboson           & $0.67\pm0.12$     & $0.97\pm0.14$        & $1.46\pm0.19$        \\
$Z$+top            & $0.62\pm0.15$     & $1.19\pm0.32$        & $1.7\pm0.5$        \\
\bottomrule
\end{tabular}
\label{tab:bkg_final}
\end{table}

The total number of background events estimated from data and MC simulation is 26.2$\pm$3.6.
Numbers of background events expected per channel 
estimated for different processes 
are shown in Table~\ref{tab:bkg_final}. 
The background estimation was cross-checked with an alternative
method, described in Refs.~\cite{atlas-singlez,smzz}, 
called the fake-factor method.
The results from this cross-check are found to be
consistent within uncertainties with those described above. 

%% file: texfiles/CrossSection_Section.tex
\section{Cross-section extraction method}
\label{sec:CrossSectionExtraction}

Two cross sections are extracted from the number of observed events. 
One is the fiducial cross section, $\sigma_{4\ell}^{\mathrm{fid}}$, in the experimental phase space 
defined by the event selection criteria and the other is the 
cross section, $\sigma_{4\ell}^{\mathrm{ext}}$, in an extended common phase space
where electrons and muons have the same geometric and kinematic acceptance.
The fiducial phase space is defined in Table~\ref{tab:FiducialRegion}.
The extended phase space for the $4\ell$  
cross-section extraction is defined by
80 $< m_{4\ell} < 1\,000~\GeV$, 
$m_{\ell^+\ell^-} > 4$~\GeV, 
$p_\mathrm{T}^{Z_{1,2}} > 2\GeV$, 
and the presence of 
four leptons each with $p_\mathrm{T} > 5 \GeV$ and $|\eta| < 2.8$.

 \newcommand{\Npred}{N_{\mathrm{pred}}}
 \newcommand{\Azz}{A_{ 4\ell}}
 \newcommand{\Czz}{C_{ 4\ell}}
 \newcommand{\NtauMC}{N_\tau^\mathrm{MC}}
 \newcommand{\NsigMC}{N_\mathrm{sig}^\mathrm{MC}}

The cross section measurement is performed using a likelihood fit described below.
For a given channel {\it i},
the observed number of events, $\Nobs^i$, 
follows a Poisson distribution,
$\mathrm{Pois}(\Nobs^i,\Npred^i)$,
the mean of which, $\Npred^i = N_\mathrm{s}^i + N_\mathrm{b}^i $, is the sum of the expectations for signal and background yields.
These yields depend
on the fiducial cross section and 
the nuisance parameters, $\vec{x}$, which represent 
the experimental and theoretical uncertainties as: 
\begin{eqnarray}
N_\mathrm{s}^i(\xsecfid,\vec{x}) &=& N_\mathrm{s}^i(\xsecfid,0) (1 + \sum_k x_k S_k^i), \\
N_\mathrm{b}^i(\vec{x}) &=& N_\mathrm{b}^i(0) (1 + \sum_k x_k B_k^i),
\end{eqnarray}
where $S_k^i$ and $B_k^i$ are the relative systematic effects on the
signal and background, respectively, due to the $k$-th source of systematic uncertainty.
The central expectation of the signal yield, 
corresponding to the systematic sources at the nominal value
(referred to as the nuisance-free expectation),
is given by :
 \begin{equation}\label{eq:Nsi}
  N_\mathrm{s}^i(\xsecfid,0) = \mathcal{L} \cdot\Czz \cdot K_{\tau} \cdot \xsecfid,
 \end{equation}
 where $\mathcal{L}$ is the integrated luminosity,
 and $\Czz$ is the ratio of the number of accepted
 signal events to the number of generated events in the fiducial phase space.
 Corrections are applied to $C_{4\ell}$ to account for measured differences
 in trigger and reconstruction efficiencies between simulated and data samples.
 The $C_{4\ell}$ values are  53.3\%, 82.2\% and 67.7\% for the $4e$, $4\mu$, 
 and $2e2\mu$ channels, respectively.
 The contribution from $\tau$-lepton decays
 is accounted for
 by a correction term  $K_{\tau} =  1+\NtauMC / \NsigMC$,
where $\NtauMC$ is the number of accepted simulated $4\ell$ events in which at least one of 
the $Z$  bosons decays into $\tau$-lepton pairs, and $\NsigMC$ is the number of accepted
simulated {\it ZZ} events with decays into electrons or muons.

\begin{table}[tb]
\centering
\caption{List of selection cuts which define the fiducial region of the cross-section measurement.
Same-flavour opposite-sign lepton pairs are denoted as SFOS, the leading lepton pair mass as $m_{12}$,
and the sub-leading lepton pair mass as $m_{34}$.
The four-momenta of all final-state photons within $\Delta R = 0.1$ of a lepton are 
added to the four-momentum of that lepton. }
\label{tab:FiducialRegion}
\vspace{0.2cm}
\begin{tabular}{ll}
\hline
\multicolumn{2}{c}{\bf Lepton selection} \\
Muons: & $\pt>6~\mathrm{GeV}, |\eta|<2.7$  \\
Electrons: & $\pt>7~\mathrm{GeV}, |\eta|<2.5$  \\
\hline
\multicolumn{2}{c}{\bf Lepton pairing} \\
Leading pair:    & SFOS lepton pair with \\
                 & smallest $|m_Z-m_{\ell\ell}|$  \\
Sub-leading pair: & The remaining SFOS \\
                 & with the largest $m_{\ell\ell}$ \\
For both pairs:    & $p_\mathrm{T}^{\ell^+\ell^-} > $ 2 GeV \\
\hline
\multicolumn{2}{c}{\bf Event selection} \\
Lepton $p_\mathrm{T}^{\ell_{1},\ell_{2},\ell_{3}}$: & $ > 20, 15, 10 (8~\rm{if}~\mu) $ \GeV   \\
Mass requirements: & $50 < m_{12} < 120$ \GeV\\
& $12 < m_{34} < 120$ \GeV  \\
Lepton separation: & $\Delta R(\ell_{i},\ell_{j}) > 0.1~(0.2)$ \\
& for same- (different-) \\
& flavour leptons \\
$J/\psi~$veto: & $m(\ell^+_{i},\ell^-_{j}) > 5$ \GeV \\
$4\ell$ mass range: & $80<m_{4\ell} < 1\,000$ GeV \\  
\hline
\end{tabular}
\end{table}

Cross-section measurements are extracted for a single channel or any combination of channels,
using a likelihood method.
The likelihood function is:
\begin{equation}\label{eq:loglike}
L(\xsecfid, \vec{x})
= \prod_{i} \mathrm{Pois}(\Nobs^i, \Npred^i(\xsecfid,\vec{x}))
 \cdot \mathrm{e}^{-\frac{\vec{x}^2}{2}},
\end{equation}
where the product runs over the channels to be considered.

For the extended phase space the likelihood function is parameterised as 
a function of the extended cross section similar to the one shown in Eq.~(\ref{eq:Nsi}) 
and multiplied by the fiducial acceptance $A_{4\ell}$, which
is the ratio of the number of events
within the fiducial phase-space region to the total number of
generated events in the extended phase space.
The fiducial acceptance $A_{4\ell}$ is evaluated using simulation 
to be 41.6\%, 50.3\%, and 42.2\%, for the $4e$, $4\mu$, and  $2e2\mu$ channels, respectively.
The differences are due to the electron and muon geometric
detection coverage. 

To find the central value of the cross section $\sigma$,
the likelihood function is maximised simultaneously with respect to
the nuisance parameters and $\sigma$.
Correlations between the signal and background systematic uncertainties 
are taken into account in the likelihood fitting procedure.

%% file: texfiles/Systematic_Section.tex
\section{Systematic uncertainties}
\label{sec:Systematic}
Systematic uncertainties on the measurement arise from uncertainties
on the integrated luminosity, the experimental calibrations of the
lepton energy and momentum, and the lepton detection efficiencies, as well as
the theoretical modelling of signal acceptance, and the background
estimation.
The overall uncertainty on the integrated luminosity is $\pm 2.8\%$,
which is derived following the same methodology as that detailed in Ref.~\cite{lumi2011}.
A summary of the relative uncertainties of 
$C_{4\ell}$, $A_{4\ell}$, and $A_{4\ell}\times C_{4\ell}$
is given in Tables~\ref{tab:acceptance_exp_8TeV} and~\ref{tab:acceptance_a}.

The effect on the expected signal event yields due to experimental
systematic uncertainties is determined from the uncertainties on
lepton energy and momentum scales and resolutions, as well as the
uncertainties on efficiencies of the lepton reconstruction and
identification.  
The major contributions come from the uncertainties
on lepton reconstruction and identification
efficiencies~\cite{atlas-muon-reco,atlas-egamma-calib,atlas-egamma-reco-id}.

The uncertainties on the signal acceptance for both $C_{4\ell}$ and
$A_{4\ell}$ include theoretical uncertainties from the choice of QCD
scales and PDF set.  The scales are varied independently from 0.5 to
2.0 times the nominal values of $\mu_\mathrm{R}$ and $\mu_\mathrm{F}$.
The PDFs uncertainties are estimated by using the
envelope~\cite{pdf4lhc} of variations of different PDF sets, CT10,
MSTW2008~\cite{bib:MSTW2008} and NNPDF2.3~\cite{NNPDF23}.

The $C_{4\ell}$ uncertainty is mostly experimental and of the order of 
2--5\%, while the $A_{4\ell}$ uncertainty is entirely theoretical and
of the order of 
3--5\%.  A range of values of the relative uncertainties on the $C_{4\ell}$
are given by 4.9\%, 1.9\%, and 2.5\% for the $4e$, $4\mu$, and $2e2\mu$,
respectively.  The uncertainties on $C_{4\ell}$ due to higher-order
corrections to the $gg$ production processes are less than 0.6\%.
This is estimated by applying an approximate NNLO $K$-factor determined
for the Higgs-boson production~\cite{kfactor-ggzz}, 
assuming that it is applicable to the
normalisation of the continuum $gg\rightarrow ZZ$ production cross
section.

Uncertainties on $C_{4\ell}$, as a function of $m_{4\ell}$ and
$p_\mathrm{T}^{4\ell}$, are also computed for the differential cross section
measurements.  In the mass region ($m_{4\ell} < $ 150~\GeV{}), the
relative uncertainties on $C_{4\ell}$ 
vary in the range of 4-9\%, 1.7-2.7\%, and
2-5\% for the $4e$, $4\mu$, and $2e2\mu$ channel, respectively.  In the
mass region $m_{4\ell} > $ 150~\GeV{}, they are almost constant as a
function of $m_{4\ell} $ and are about 4\%, 1.8\%, and 3\% for the $4e$,
$4\mu$, and $2e2\mu$ channel, respectively.

The relative uncertainties on $A_{4\ell}$ are 1.2\%, 1.0\%, and
1.6\% for the $4e$, $4\mu$, and $2e2\mu$ channel, respectively, evaluated
by comparing \progname{POWHEG-BOX} and \progname{MCFM} MC samples with
the same approach for the QCD scales and the PDF uncertainties as
described earlier.  The QCD scale uncertainties do not change when
going from NLO to NNLO for the signal normalisation for the $q\bar q
\rightarrow 4\ell$ events~\cite{nnlocorr}.  An additional uncertainty
(3-4\%) is included in the $A_{4\ell}$ uncertainty estimate to account
for the uncertainty of the Higgs-boson NNLO $K$-factor normalisation
correction of the non-resonant $4\ell$ signal from gluon fusion
(labelled ``extra gg corrections'' in
Tables~\ref{tab:acceptance_exp_8TeV} and~\ref{tab:acceptance_a}).

\begin{table} [htb]
\centering
\caption{
  The combined relative uncertainties on the efficiency correction factor $C_{4\ell}$,
  evaluated as the sum in quadrature of the uncertainties from different sources,
  including electron and muon identification and 
  theoretical uncertainties due to PDFs, QCD scales, and parton shower modelling.
  Extra uncertainties due to higher-order corrections for the $gg$ process
  (NNLO $K$-factors for Higgs-boson production applied to the inclusive $gg$ process)
  are also given.
}
\label{tab:acceptance_exp_8TeV}
\begin{tabular}{lccc}
\toprule
Sources                     & \multicolumn{3}{c}{$\Delta C_{4\ell} / C_{4\ell}$}             \\
                                   & $4e$  & $4\mu$ & $2e 2\mu$ \\

\midrule
\addlinespace
Experimental ($e$)     &  4.8\% & - & 2.3\% \\
\addlinespace
Experimental ($\mu$)     & - & 1.8\% & 0.9\% \\
\addlinespace
Theoretical              & 0.1\% & 0.1\%  & 0.2\%   \\
\addlinespace
Extra $gg$ corrections           & 0.6\% & 0.2\%  & 0.3\%   \\
\midrule
Combined uncertainty               & 4.9\% & 1.9\%  & 2.5\%   \\
\bottomrule
\end{tabular}
\end{table}

\begin{table} [htb]
\centering
\caption{
  Theoretical relative uncertainties on the fiducial acceptance 
  $A_{4\ell}$ and $A_{4\ell}\times C_{4\ell}$ 
  due to PDFs, QCD scales and parton shower modelling.
  Extra uncertainties due to higher-order corrections for the $gg$ process
  (NNLO $K$-factors for Higgs-boson production applied to the inclusive $gg$ process)
  are also given.
}
\label{tab:acceptance_a}
\begin{tabular}{lccc}
\toprule
                          &  \multicolumn{3}{c}{$\Delta A_{4\ell} / A_{4\ell}$}             \\

Sources                   &  $4e$     & $4\mu$ &  $2e 2\mu$ \\
\midrule
Theoretical   & 1.2\%     & 1.0\%  &   1.6\%    \\
\addlinespace
Extra $gg$ corrections  & 4.0\%     & 3.0\%  &   3.9\%   \\
\midrule
                          &  \multicolumn{3}{c}{$\Delta (A_{4\ell} \times C_{4\ell}) / (A_{4\ell} \times C_{4\ell})$}             \\
\midrule
Theoretical            & 1.4\% & 1.1\% & 1.7\%    \\
\addlinespace
Extra $gg$ corrections           & 4.6\% & 3.2\% & 4.2\%   \\

\bottomrule
\end{tabular}
\end{table}

The overall uncertainty on the background estimation is $\pm$14\%. The contributions
from different sources and channels are given in Table~\ref{tab:bkg_final}.

%% file: texfiles/Results_section.tex
\section{Results}
\label{sec:Results}

\subsection{Cross-section measurements}
\label{sec:CrossSection}

\begin{figure}[htb]
\centering
\includegraphics[width=0.8\linewidth]{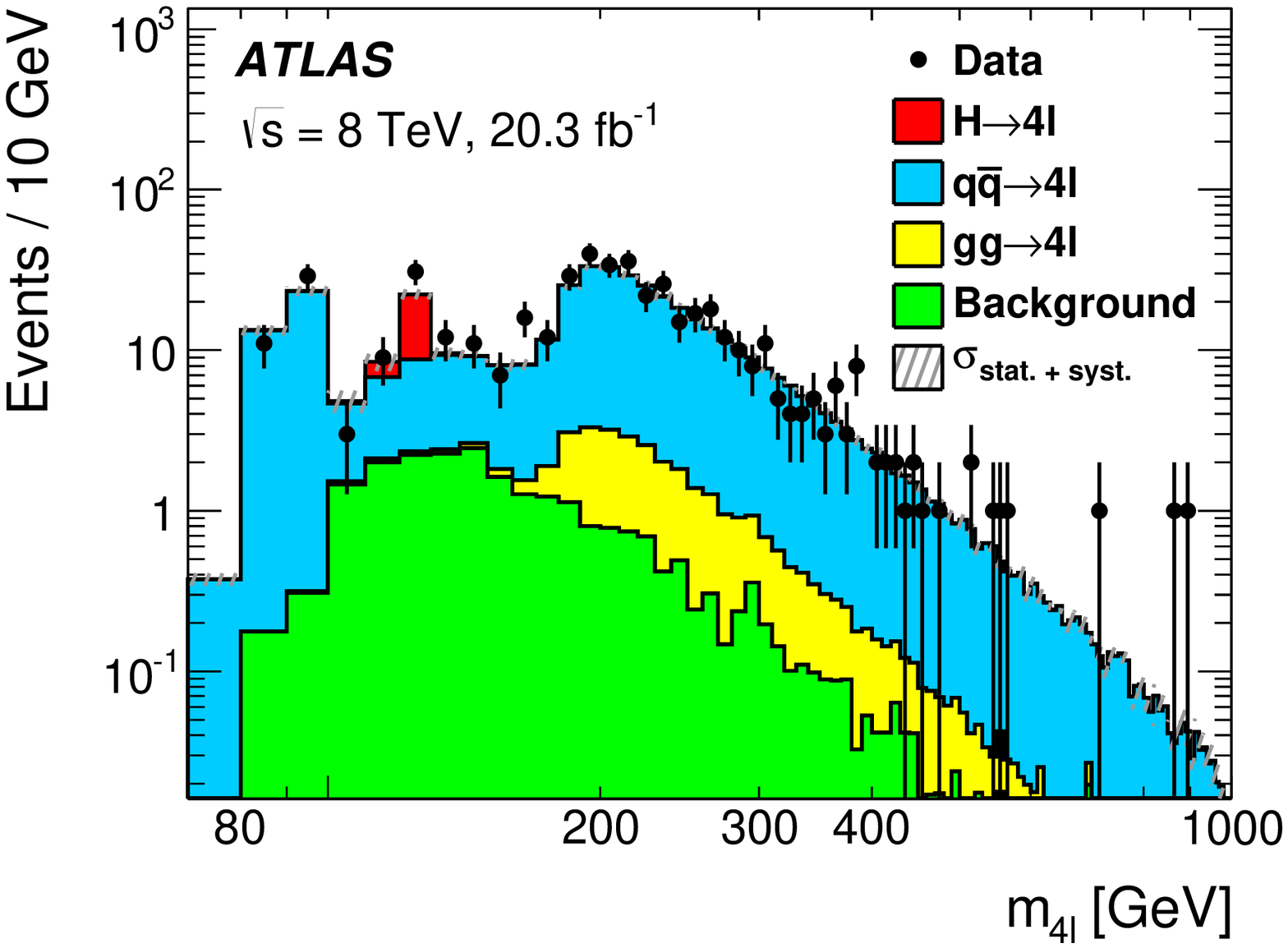}
\includegraphics[width=0.8\linewidth]{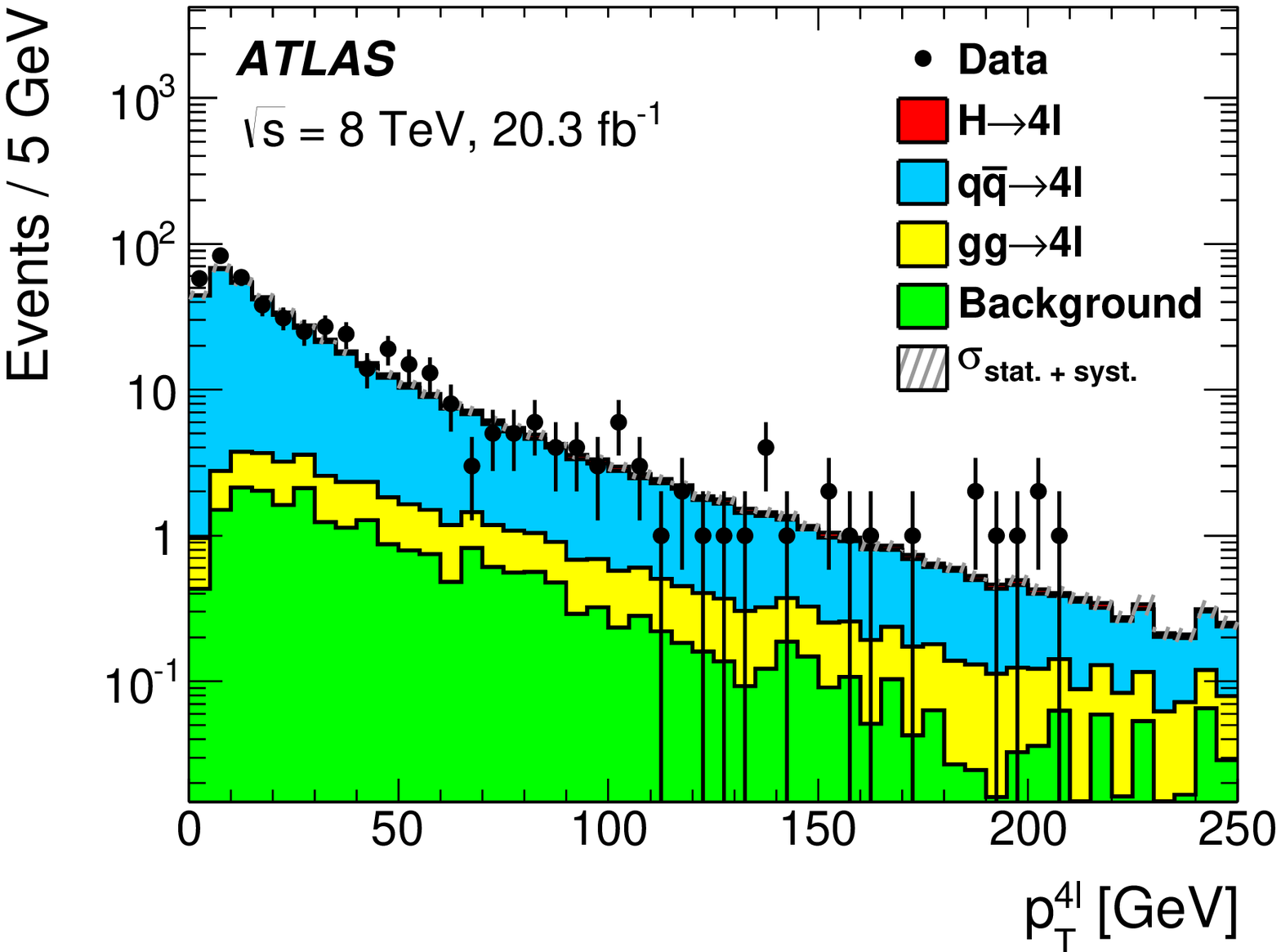}
\caption
{Data and MC prediction comparison for selected events
as a function of the invariant mass $m_{4\ell}$ (top) 
and the transverse momentum $p_\mathrm{T}^{4\ell}$ (bottom) of the four-lepton system. 
The solid colours show the expected contributions from signal and background
and the black points represent data with statistical error bars.
}
\label{fig:4l_massZZpt_logx}
\end{figure}

The numbers of expected and observed events after applying all selection criteria are shown in 
Table~\ref{tab:summary2}. 
A total of 476 candidate events is observed with a background
expectation of $26.2 \pm 3.6$ events. 
The observed and predicted
$m_{4\ell}$ and $p_\mathrm{T}^{4\ell}$ 
distributions 
for the selected events are shown in Figure~\ref{fig:4l_massZZpt_logx}.

The measured cross sections in the fiducial and extended phase space for 
different $4\ell$ channels 
are summarised 
in Table~\ref{table:xsection} 
and compared to the SM predicted cross sections.
The combined $4\ell$ cross section in the extended phase space
is found to be 
$73 \pm 4$~(stat.)~$\pm 4$~(syst.)~$\pm 2$~(lumi.)~fb,
compared to a SM prediction of  $ 65 \pm 4 $~fb.
One should note that the cross section for 
non-resonant {\it ZZ} production from the $gg$-induced signal
is only calculated at LO approximation, which could be significantly underestimated.

\begin{table*}
  \centering
  \caption{
   Summary of the observed and predicted numbers of $4\ell$ events in different $4\ell$ channels.
   $N^{\mathrm{Data}}$ denotes the selected number of data candidates.
   $N^{\mathrm{Total}}_{\mathrm{expected}}$ 
   denotes the total predicted number of events (including $\tau$ contributions)
   for signal plus background.
   $N^{\mathrm{signal}}_{\mathrm{non-}gg}$ and the 
   $N^{\mathrm{signal}}_{gg}$ denote the predicted non-$gg$ signal and
   the $gg$ signal (no NNLO $K$-factor has been applied),
   respectively.
   $N^{\mathrm{MC}}_\tau$ denotes the $\tau$ contributions.
   $N_{\mathrm{bkg}}$ denotes the total estimated number of background events (from data and MC simulation).
   The listed uncertainties of the expected number of signal events
   include statistical and experimental systematic uncertainties.
  }
  \label{tab:summary2}
  \begin{tabular}{ccccccc}
    \toprule
Channel   & $N^{\mathrm{Data}}$ & $N^{\mathrm{Total}}_{\mathrm{expected}}$ & 
            $N^{\mathrm{signal}}_{\mathrm{non-}gg}$ & $N^{\mathrm{signal}}_{gg}$ & 
            $N^{\mathrm{MC}}_\tau$ & $N_{\mathrm{bkg}}$  \\
    \midrule
    \addlinespace
    $4e$     & 85  & $80    \pm 4  $ & $68.4  \pm 3.4$ & $6.24  \pm 0.31$ & $1.28 \pm 0.06$ & $3.6   \pm 0.5$  \\
    $4\mu$   & 156 & $150.2 \pm 2.9$ & $128.2 \pm 2.5$ & $11.00 \pm 0.21$ & $2.18 \pm 0.09$ & $9.0   \pm 1.5$  \\
    $2e2\mu$ & 235 & $205   \pm 5  $ & $172   \pm 5  $ & $16.0  \pm 0.4 $ & $3.08 \pm 0.13$ & $13.6  \pm 2.1$  \\
    \midrule
    Total    & 476 & $435   \pm 9  $ & $369   \pm 9  $ & $33.3  \pm 0.8 $ & $6.54 \pm 0.14$ & $26.2  \pm 3.6$  \\
    \bottomrule
  \end{tabular}
\end{table*}

\begin{table*}[htb]
\centering
\caption{
         Measured cross sections in the fiducial phase space ($\sigma^\mathrm{fid}$) 
         and extended phase space ($\sigma^\mathrm{ext}$), 
         compared to their SM predictions (calculations described in Section~\ref{sec:Theory}). 
         One should note that the non-resonant $gg$-induced signal cross section
         is only calculated at LO approximation. 
}
{\small
\begin{tabular}{lcccc}
    \toprule
$4\ell$ & Measured $\sigma^\mathrm{fid}$ [fb] & SM $\sigma^\mathrm{fid}$ [fb]
        & Measured $\sigma^\mathrm{ext}$ [fb] & SM $\sigma^\mathrm{ext}$ [fb] \\
    \midrule
    \addlinespace
 $4e$  & 7.4$^{+0.9}_{-0.8}$ (stat) $^{+0.4}_{-0.3}$ (syst) $^{+0.2}_{-0.2}$ (lumi) & $6.9 \pm 0.4$ 
       & 17.8$^{+2.1}_{-2.0}$ (stat) $^{+1.5}_{-1.1}$ (syst) $^{+0.5}_{-0.5}$ (lumi)& $16.4\pm 1.0$\\
    \addlinespace
 $4\mu$& 8.7$^{+0.8}_{-0.7}$ (stat) $^{+0.2}_{-0.2}$ (syst) $^{+0.3}_{-0.2}$ (lumi) & $8.3 \pm 0.5$
       &17.3$^{+1.5}_{-1.4}$ (stat) $^{+0.9}_{-0.7}$ (syst) $^{+0.5}_{-0.5}$ (lumi) & $16.4\pm 1.0$\\
    \addlinespace
$2e2\mu$& 15.9$^{+1.1}_{-1.1}$ (stat) $^{+0.5}_{-0.4}$ (syst) $^{+0.5}_{-0.4}$ (lumi)& $ 13.7 \pm 0.9$ 
        &37.7$^{+2.7}_{-2.6}$ (stat) $^{+2.5}_{-2.0}$ (syst) $^{+1.1}_{-1.1}$ (lumi) & $ 32.1 \pm 2.0 $\\
    \midrule
    \addlinespace
Total   &                                                                               &
        & 73$^{+4}_{-4}$ (stat) $^{+4}_{-4}$ (syst) $^{+2}_{-2}$ (lumi)              & $ 65 \pm 4 $\\
    \bottomrule
\end{tabular}
}
\label{table:xsection}
\end{table*}

%% file: texfiles/Unfolding_Section.tex
\subsection{Differential cross-section measurement}
\label{sec:Unfolding}
The measurement of the differential cross-section is performed in the fiducial
phase space defined in Table \ref{tab:FiducialRegion}. 
The events from all three $4\ell$ channels are combined into
a common sample for the unfolding procedure.
The unfolding is done as 
a function of the two kinematic variables $m_{4\ell}$ and $p_\mathrm{T}^{4\ell}$. 
The $m_{4\ell}$ 
spectrum is  essential for the study of the different production mechanisms, 
while the $p_\mathrm{T}^{4\ell}$ spectrum is sensitive to 
higher-order QCD corrections and to QCD resummation effects at 
small $p_\mathrm{T}^{4\ell}$~\cite{resummation}.
The high-$p_\mathrm{T}^{4\ell}$ region is sensitive to 
top-loop effects in $gg\rightarrow H$ production as well as to
anomalous triple-boson couplings.

The iterative Bayesian unfolding~\cite{iterDAgostini} is applied here.
In the unfolding of binned data, the effects of the experimental
acceptance and resolution are expressed in terms of a response matrix,
where each element corresponds to the probability of an event in the $i$-th generator level bin
being reconstructed in the $j$-th measurement bin.
The response matrix
is combined with the measured spectrum to form a likelihood, which is
then multiplied by a prior distribution to produce the posterior
probability of the true spectrum. The SM prediction is used as the
initial prior, and once the posterior probability is obtained, it is
used as the prior for the next iteration. 
The spectrum becomes insensitive to the initial prior after a few
iterations.
The differences between successive iterations are used to estimate the
stability of the unfolding method. In this analysis four iterations
are performed.

\begin{figure*}[tb]
\centering
\includegraphics[width=0.48\linewidth]{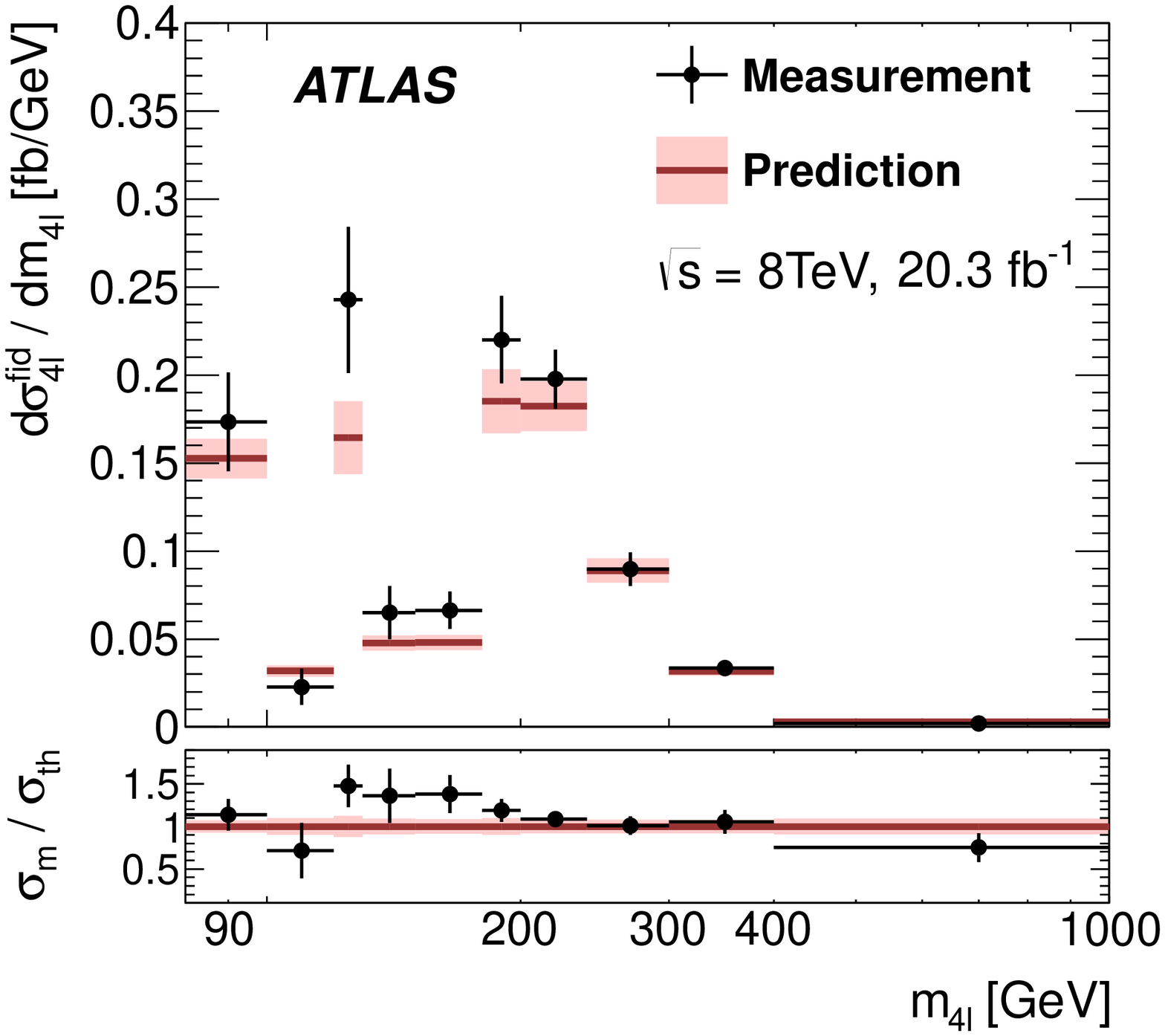}
\includegraphics[width=0.48\linewidth]{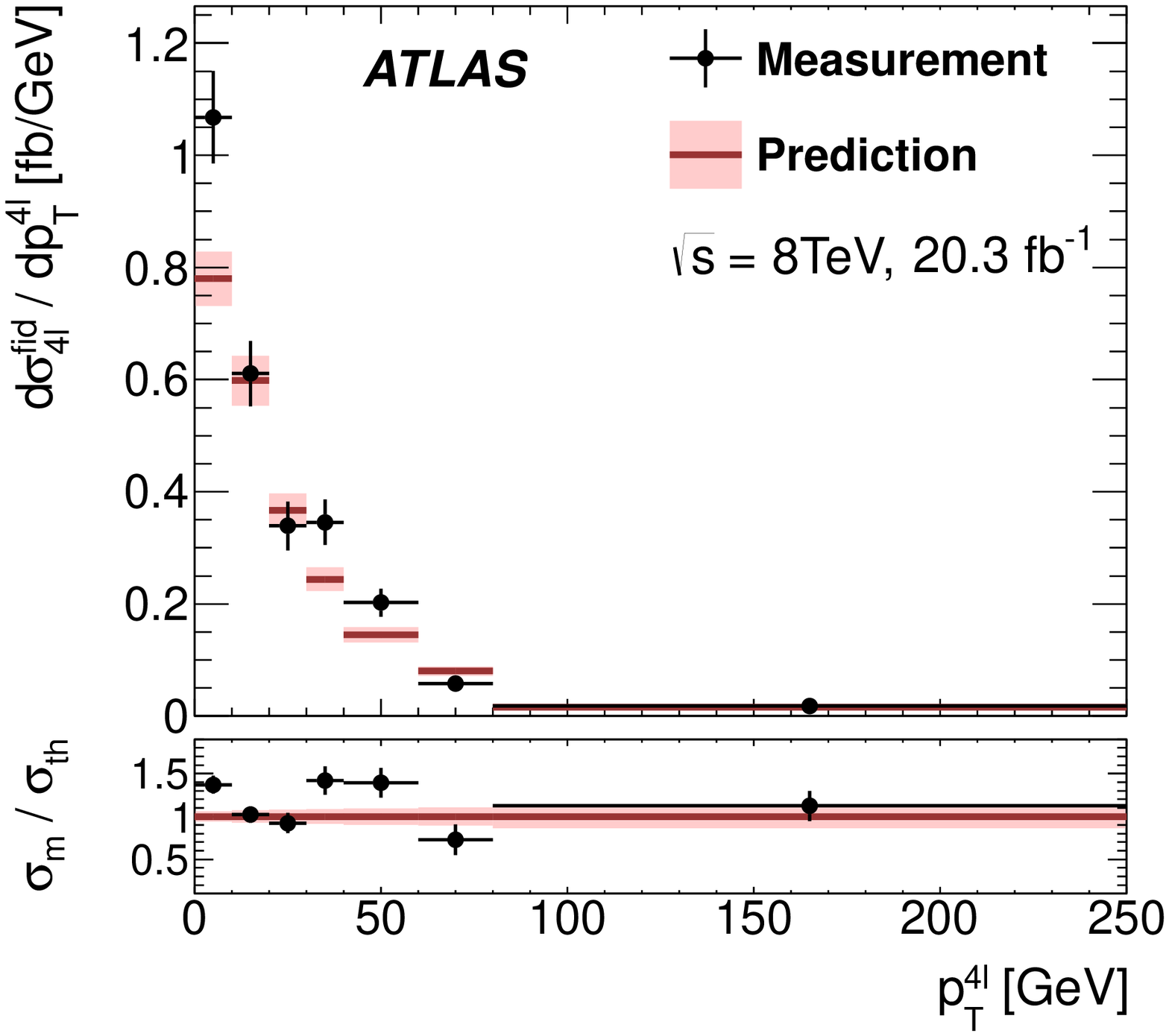}
\caption{
The measured differential cross-section distributions 
(the black points) 
of $m_{4\ell}$ (left) and $p_\mathrm{T}^{4\ell}$ (right),
unfolded into the fiducial phase space,
and compared to theory predictions (red histogram).
The combined statistical and systematic uncertainties of the measurements
are shown as the error bars of the unfolded spectra.
The theoretical predictions are the sum of the differential cross sections
of the $q\bar q\rightarrow 4\ell$ and $gg\rightarrow 4\ell$ processes,
where the LO cross sections are used for the non-resonant $gg$-induced signals,
and the cross sections of the on-shell Higgs boson and the $q\bar q$ production processes  
are corrected with the NNLO $K$-factors for the $m_{4\ell}$ spectrum;
except for the $p_\mathrm{T}^{4\ell}$ where only the NLO and LO predictions 
are used for the $q\bar q$ and the $gg$ processes, respectively.
The total theoretical uncertainties are shown as error bands evaluated by
the sum in quadrature of the contributions from
parton showers, QCD scales, PDF sets, and electroweak corrections. 
}
\label{fig:unfold_fv}
\end{figure*}

The unfolded distributions are shown in Figure~\ref{fig:unfold_fv}, 
where the differential cross section 
is presented as a function of $m_{4\ell}$ and $p_\mathrm{T}^{4\ell}$ 
and compared to theory predictions.
The data points shown in the figures are the measurements
with combined statistical and systematic uncertainties.
The theoretical predictions are the sum of the differential cross sections
of the $q\bar q\rightarrow 4\ell$ and $gg\rightarrow 4\ell$ processes.
The LO cross sections are used for the non-resonant $gg$-induced signals.
The cross sections of the on-shell Higgs boson are normalised to include the NNLO QCD and NLO EW effects as summarised in Ref.~\cite{yellowreport}.
The $q\bar q$ production processes are corrected with NNLO QCD and the NLO EW $K$-factors for the $m_{4\ell}$ spectrum for
$m_{4\ell} > 2\times m_Z$.
For the $p_\mathrm{T}^{4\ell}$ spectrum, the $q\bar q$ signal prediction is calculated 
by \progname{POWHEG-BOX} at NLO.

The uncertainties on the differential cross-section measurements are dominated by 
the statistical uncertainties of the data. 
For example, in the $m_{4\ell}$ regions 
between the Z and Higgs boson peaks and 
between the Higgs-boson mass $m_H$ and $m_{4\ell} = 180$ GeV,  
the statistical uncertainties 
are of the order of 45\% and 20\%, respectively.
In the high-mass region ($m_{4\ell} > 180$ GeV) they are of the order of 10\%.
Furthermore, one should note that
the NNLO QCD corrections are not available for the $q\bar q \rightarrow 4\ell$ production calculation 
for the mass region $m_{4\ell} < 2\times m_{Z}$.

In the $m_{4\ell}$ bin of 120--130 GeV, which is dominated by the resonant Higgs-boson contribution, 
the ratio of data to the MC prediction is 
compatible with the ATLAS measurement~\cite{atlas-higgs4lRun1}
of the Higgs-boson signal strength of $\mu_H = 1.44^{+0.40}_{-0.33}$. 
The data points in the $m_{4\ell}$ spectrum between 140 to 180 GeV are slightly more than 1$\sigma$
above the theoretical predictions, 
where the NNLO QCD correction is not yet available.
Some discrepancy is also observed 
in the lowest bin and in the region between 30--50~GeV of the $p_\mathrm{T}^{4\ell}$ spectrum.

%% file: texfiles/ggExtraction_Section.tex
\subsection{Extraction of the $gg$ signal contribution in the $m_{4\ell} > 180$~GeV region}
\label{sec:ggExtraction}

The extraction of the
signal strength of the non-resonant 
$gg\rightarrow 4\ell$ production is performed in the high-mass
region, $m_{4\ell} > 180$~GeV,
where this production mode is dominated by the continuum $gg\rightarrow ZZ$ process through 
a quark-box diagram intermediate state (see Figure~\ref{fig:qqZZ-feynman}(d)). 
Additional contributions come from 
the off-shell Higgs-boson production and the interference between Higgs boson  
and continuum $ZZ$ production.

The $m_{4\ell}$ spectrum is chosen as the discriminant 
to extract the $gg$ signal strength with respect to the LO $gg$ prediction:
$\mu_{gg} = \sigma(\mathrm{data})/\sigma(\mathrm{LO})$.

\begin{figure}[bt]
\centering
\includegraphics[width=0.95\linewidth]{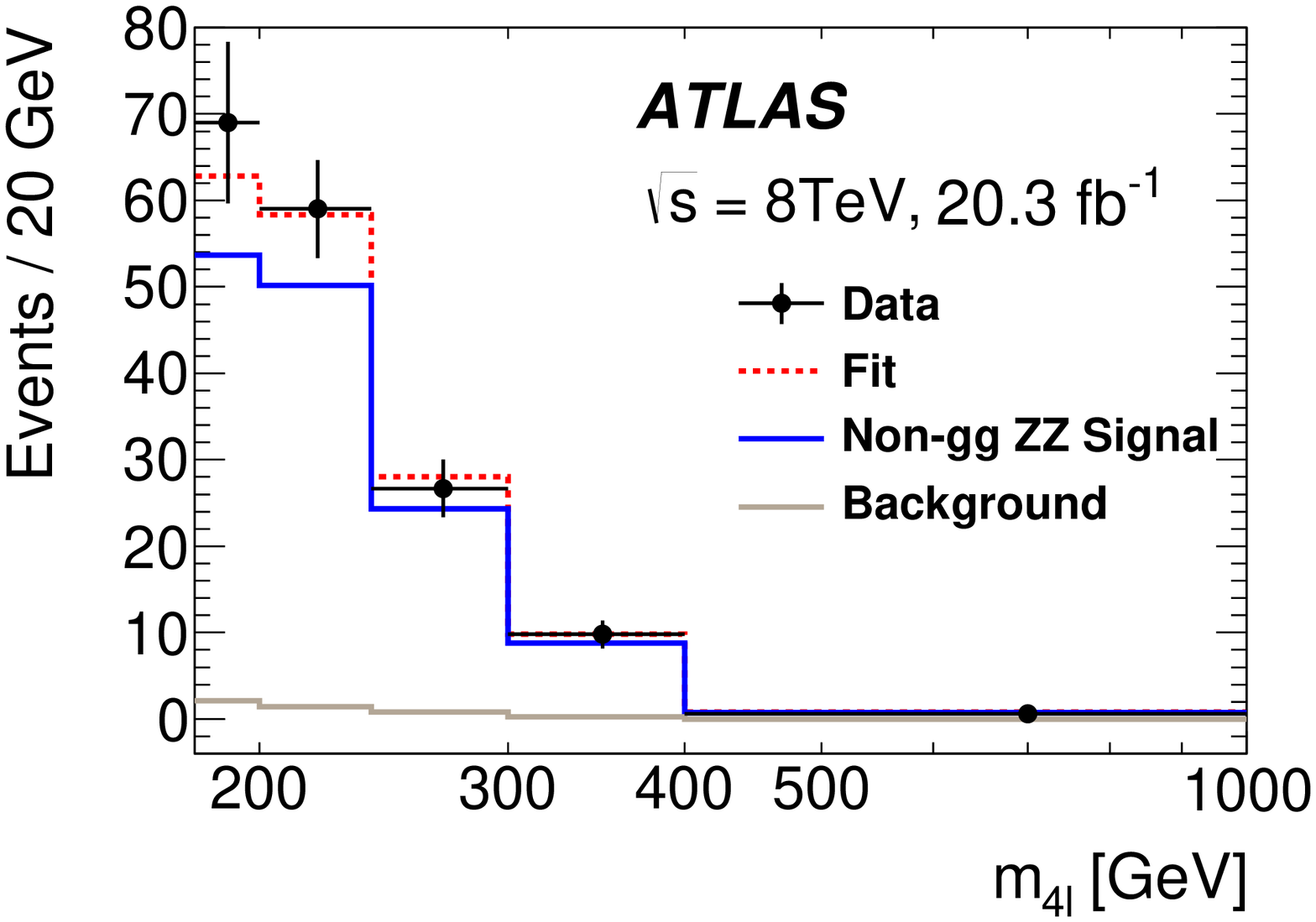}
\caption{Comparison of the $m_{4\ell}$ spectra 
         between the data (black points with error bars) 
         and the prediction (red histogram) after the 
         likelihood fit of $\mu_{gg}$.
         The non-$gg$ signal from the theoretical prediction (blue histogram)
         and the background (brown histogram) are also shown.
         The $gg$ contribution is the difference between data and the
         sum of the non-$gg$ signal and the background.}
\label{fig:app_ggfit}
\end{figure}

The contribution of the $q \bar q\rightarrow ZZ$ production is constrained 
to the best theory knowledge
(which accounts for QCD NNLO and EW NLO $m_{4l}$-dependent corrections)
and  $\mu_{gg}$ is extracted from a likelihood fit using the
reconstructed $m_{4\ell}$ distributions.
The experimental uncertainties are treated 
as fully correlated between $q \bar q$ and $gg$ processes.
The theoretical uncertainties,
including the uncertainties on the normalisation of the $q\bar q\rightarrow ZZ \rightarrow 4\ell$,
the shapes of $4\ell$ spectra from both the $q\bar q$ and $gg$ initial states,
and the acceptance,
are taken into account. 
The $m_{4\ell}$ distribution of the data, the fit, 
the expectation from non-$gg$ signal processes and the background
are shown in Figure~\ref{fig:app_ggfit}.
The fit result is 
$
\mu_{gg}~=~2.4 \pm 1.0 (\mathrm{stat.}) \pm 0.5 (\mathrm{syst.}) \pm 0.8 (\mathrm{theory}).
$
This result corresponds to a $gg$-initiated cross section of 3.1 fb,
which has the same relative uncertainties as $\mu_{gg}$ itself in the
inclusive fiducial volume as defined in  Table \ref{tab:FiducialRegion} with the
additional requirement of $m_{4\ell} > $~180~\GeV{}.
The largest uncertainty is statistical. 
The theoretical uncertainty is 
mainly due to the normalisation uncertainty of the $q\bar q\rightarrow ZZ$ process.

The theoretical estimate of 
$m_{4\ell}$-dependent $K$-factor for off-shell Higgs boson production 
given in Ref.~\cite{kfactor-ggzz} is
in a range of 2.7--3.1 (with CT10NNLO PDF) and that given in
Ref.~\cite{kfactor-ggzzinterference} for the interference term 
is 2.05-2.45. 
These theoretical studies confirm that the gluon soft-collinear approximation 
predicts similar $K$-factors for off-shell Higgs-boson and interference, 
hence supporting the assumption of a similar $K$-factor for the continuum $ZZ$ production.
These theoretical calculated $K$-factors
are compatible with
the result obtained by this analysis, 
where the $gg$-initiated $4\ell$ events are produced predominantly from the 
continuum $ZZ$ production.

Applying the higher-order corrections to both the cross section of the off-shell
Higgs-boson production and the contribution of the interference term, while keeping the LO
cross section for the continuum $gg\rightarrow ZZ$ production, the change of the
$\mu_{gg}$ fit result
is negligible (approximately, $\Delta \mu_{gg} = 0.01$).

%% file: texfiles/SummaryConclusions_Section.tex
\section{Conclusion}
\label{sec:SummaryConclusion}

The measurement of four-lepton production in proton--proton collisions at $\sqrt{s}=8$~\TeV{} 
is presented using data corresponding to an integrated luminosity of 20.3 fb$^{-1}$  
collected with the ATLAS detector at the LHC. 
In total, 476 $4\ell $ candidate events are observed, 
with a background expectation of $26.2 \pm 3.6$ events, 
in the four-lepton invariant mass range between 80 and 1\,000 GeV.
The $4\ell$ production cross sections are determined in both
fiducial and extended phase spaces.  
The measured cross section in the extended phase space, 
defined by
80 $< m_{4\ell} < 1\,000~\GeV$, 
$m_{\ell^+\ell^-} > 4$~\GeV, 
$p_\mathrm{T}^{Z_{1,2}} > 2\GeV$, 
four leptons each with $p_\mathrm{T} > 5 \GeV$ and $|\eta| < 2.8$,
is found to be 
$73 \pm 4$~(stat.)~$\pm 4$~(syst.)~$\pm 2$~(lumi.)~fb,
and  is compared to a SM prediction of  $  65 \pm 4 $~fb.
The measurements  of the $4\ell$  differential cross sections 
are performed by unfolding the  $m_{4\ell}$ and the $p_\mathrm{T}^{4\ell}$ spectra. 
In the mass range above $180$~\GeV{},
assuming the theoretical constraint on the $q\bar q$ production cross section
calculated with perturbative NNLO QCD and NLO electroweak corrections,
the signal strength of the gluon-fusion component
with respect to the LO prediction
is determined to be
$\mu_{gg} =2.4 \pm 1.0 ~(\mathrm{stat.}) \pm 0.5 ~(\mathrm{syst.}) \pm 0.8 ~(\mathrm{theory})$.

%% file: acknowledgements/Acknowledgements.tex

\vspace{0.3cm}
\noindent {\bf  Acknowledgements}
\vspace{0.3cm}

We thank CERN for the very successful operation of the LHC, as well as the
support staff from our institutions without whom ATLAS could not be
operated efficiently.

We acknowledge the support of ANPCyT, Argentina; YerPhI, Armenia; ARC, Australia; BMWFW and FWF, Austria; ANAS, Azerbaijan; SSTC, Belarus; CNPq and FAPESP, Brazil; NSERC, NRC and CFI, Canada; CERN; CONICYT, Chile; CAS, MOST and NSFC, China; COLCIENCIAS, Colombia; MSMT CR, MPO CR and VSC CR, Czech Republic; DNRF, DNSRC and Lundbeck Foundation, Denmark; IN2P3-CNRS, CEA-DSM/IRFU, France; GNSF, Georgia; BMBF, HGF, and MPG, Germany; GSRT, Greece; RGC, Hong Kong SAR, China; ISF, I-CORE and Benoziyo Center, Israel; INFN, Italy; MEXT and JSPS, Japan; CNRST, Morocco; FOM and NWO, Netherlands; RCN, Norway; MNiSW and NCN, Poland; FCT, Portugal; MNE/IFA, Romania; MES of Russia and NRC KI, Russian Federation; JINR; MESTD, Serbia; MSSR, Slovakia; ARRS and MIZ\v{S}, Slovenia; DST/NRF, South Africa; MINECO, Spain; SRC and Wallenberg Foundation, Sweden; SERI, SNSF and Cantons of Bern and Geneva, Switzerland; MOST, Taiwan; TAEK, Turkey; STFC, United Kingdom; DOE and NSF, United States of America. In addition, individual groups and members have received support from BCKDF, the Canada Council, CANARIE, CRC, Compute Canada, FQRNT, and the Ontario Innovation Trust, Canada; EPLANET, ERC, FP7, Horizon 2020 and Marie Skłodowska-Curie Actions, European Union; Investissements d'Avenir Labex and Idex, ANR, Region Auvergne and Fondation Partager le Savoir, France; DFG and AvH Foundation, Germany; Herakleitos, Thales and Aristeia programmes co-financed by EU-ESF and the Greek NSRF; BSF, GIF and Minerva, Israel; BRF, Norway; the Royal Society and Leverhulme Trust, United Kingdom.

The crucial computing support from all WLCG partners is acknowledged
gratefully, in particular from CERN and the ATLAS Tier-1 facilities at
TRIUMF (Canada), NDGF (Denmark, Norway, Sweden), CC-IN2P3 (France),
KIT/GridKA (Germany), INFN-CNAF (Italy), NL-T1 (Netherlands), PIC (Spain),
ASGC (Taiwan), RAL (UK) and BNL (USA) and in the Tier-2 facilities
worldwide.

%% file: atlas_authlist.tex
\begin{flushleft}
{\Large The ATLAS Collaboration}

\bigskip

G.~Aad$^{\rm 85}$,
B.~Abbott$^{\rm 113}$,
J.~Abdallah$^{\rm 151}$,
O.~Abdinov$^{\rm 11}$,
R.~Aben$^{\rm 107}$,
M.~Abolins$^{\rm 90}$,
O.S.~AbouZeid$^{\rm 158}$,
H.~Abramowicz$^{\rm 153}$,
H.~Abreu$^{\rm 152}$,
R.~Abreu$^{\rm 116}$
% ...
\bigskip
\\
% ...
$^{11}$ Institute of Physics, Azerbaijan Academy of Sciences, Baku, Azerbaijan\\
% ...
$^{85}$ CPPM, Aix-Marseille Universit{\'e} and CNRS/IN2P3, Marseille, France\\
% ...
$^{90}$ Department of Physics and Astronomy, Michigan State University, East Lansing MI, United States of America\\
% ...
$^{107}$ Nikhef National Institute for Subatomic Physics and University of Amsterdam, Amsterdam, Netherlands\\
% ...
$^{113}$ Homer L. Dodge Department of Physics and Astronomy, University of Oklahoma, Norman OK, United States of America\\
% ...
$^{116}$ Center for High Energy Physics, University of Oregon, Eugene OR, United States of America\\
% ...
$^{151}$ Institute of Physics, Academia Sinica, Taipei, Taiwan\\
$^{152}$ Department of Physics, Technion: Israel Institute of Technology, Haifa, Israel\\
$^{153}$ Raymond and Beverly Sackler School of Physics and Astronomy, Tel Aviv University, Tel Aviv, Israel\\
% ...
$^{158}$ Department of Physics, University of Toronto, Toronto ON, Canada\\
% ...
\end{flushleft}
